\documentclass[fleqn,twoside]{article}
\usepackage[headings]{espcrc2}

\readRCS
$Id: espcrc2.tex,v 1.2 2004/02/24 11:22:11 spepping Exp $
\usepackage{graphicx}
\usepackage[figuresright]{rotating}

\def\ebeam{E_{\mathrm{beam}}}
\def\roots{\sqrt s}
\def\myratio{{\cal R}_{\tau\tau}^{\Upsilon} }
\def\myratiooff{{\cal R}_{\tau\tau}^{\mathrm{Off}}}
\def\myratiooffexp{{\cal R}_{\tau\tau}^{\mathrm{Off} \, {\mathrm{theory}}}}

\def\etal{{\it et~al.}}

\title{Universality and Lepton Flavor Violation
 in $\Upsilon$ Decays at CLEO }

\author{J.E. Duboscq\address[Wilson]{Wilson Laboratory, Cornell University,
 Ithaca, NY, 14853, USA}, for the CLEO Collaboration
       \thanks{I would like to thank A.Lusiani and the staff of Tau06
 for a wonderfully stimulating conference.}
}

\runtitle{Universality and Lepton Flavor Violation
 in $\Upsilon$ Decays at CLEO }
\runauthor{J.E. Duboscq}

\begin{document}

\begin{abstract}
 I present two analyses done with the CLEOIII detector at CESR. The first
 is a search for the LFV decay $\Upsilon\to\mu\tau$  with preliminary
 results, and the second
 constrains Lepton Universality in $\Upsilon\to\tau\tau$ relative
 to $\Upsilon\to\mu\mu$. This second analysis, whose results are final,
  has dramatically 
 improved uncertainties relative to other such measurements, and
 is also the first observation of the decay $\Upsilon(3S)\to\tau\tau$.
 A limit is also set on the involvement of a CP odd Higgs in the decay
 of the $\Upsilon(1S)$ to $\tau$ pairs.
\vspace{1pc}
\end{abstract}

\maketitle

\section{Introduction}
In this report, I present results on two analyses done with data
 taken using the CLEOIII detector on the CESR storage ring by the
 CLEO Collaboration.
 The first analysis is a search for the Lepton Flavor Violating (LFV) 
 decay $\Upsilon(nS) \to \mu\tau$,
 ${\rm n=1,2,3}$ - all results from this analysis are preliminary.
 The second analysis tests Lepton Universality in the
 decays $\Upsilon \to \mu\mu, \tau\tau$,  ${\rm n=1,2,3}$ - all 
 results from this analysis are final.
   
 The CLEOIII detector is a modern almost hermetic detector, sitting
 on the CESR beamline, where positrons and electrons collide with an
 energy $\approx 5 {\rm \,GeV}$. The detector included a silicon strip 
 detector at its center, surrounded by a wire chamber, which provided
 tracking and dE/dx information, a Cesium Iodide calorimeter, and
 a Ring Imaging Cerenkov Detector all in a 1.5 Tesla magnetic field.
 Outside of this were a series of muon chambers. Further details of
 the detector can be found in \cite{CLEOdet}.

 The data used in this analysis comprise the $\Upsilon(nS)$,  ${\rm n=1,2,3}$
 datasets, along with a portion of the $\Upsilon(4S)$ data for validation
 purposes. These datasets include data taken at the peaks of the resonances,
 as well as smaller datasets taken some $30 - 40 {\rm MeV}$ below
 the resonances. The total sample includes approximately $20 \times 10^6$
 $\Upsilon(1S)$, $10 \times 10^6$ $\Upsilon(2S)$ and 
 $5 \times 10^6$ $\Upsilon(3S)$ decays.

\section{Search for LFV in $\Upsilon \to \mu\tau$ }
Lepton Flavor Violation (LFV) might provide a key to understanding 
 Lepton Number Violation, as well as Baryon Number Violation, and hence
 might be helpful in understanding the matter/anti-matter asymmetry in 
 the Universe. 
LFV decays in the $\Upsilon$ sector can be related to LFV decays of the
 $\tau$ lepton by a simple reordering of input and output lines in
 the Feynman diagram for the process $e^+e^-\to \gamma* \to \Upsilon*
  \to \tau \mu $. In this way, ~\cite{taulfv} have show that a result
 of $B(\tau \to\ 3\mu) < 10^{-6} $ implies that $B(\Upsilon \to
 \mu \tau) < 10^{-2}$. 

LFV decays of the $\Upsilon$ to lepton pairs might also be expected 
 in the presence of SUSY loops ~\cite{SUSY}, or other exotic physics such as 
 leptoquarks and theories with more than one Z boson.

A general ansatz for talking about LFV decays of the $\Upsilon$ involves
 a generic four fermion interaction vertex between two b quarks, a muon and
 a $\tau$. With the coupling constant at the interaction denoted $\alpha_N$ 
 and the relevant new physics mass scale denoted $\Lambda$, one finds
 ~\cite{fourfemion}:
 $$ {B(\Upsilon \to \mu\tau)\over{B(\Upsilon \to \tau\tau)}}
 \propto (\alpha_N/\alpha)^2 (M_\Upsilon/\Lambda)^4 $$.

The CLEO analysis presented here is a search for 
 $\Upsilon(nS)\to \mu\tau,\, \tau \to e\nu\nu$, n=1,2,3. The final
 observed state is a two track event with a muon and a electron
 and missing energy. The muon would have an energy a little 
 below the beam energy, while the electron energy spectrum
 would be approximately that expected for electrons in normal
 $\tau $ pair events.

The fit to the data is performed using an extended maximum
 likelihood function, composed of a product PDF, summed over
 expect signal LFV shapes, direct $ee\to \tau\tau$  shapes, $ee\to\mu\mu\gamma$
 and $ee\to\mu\mu$ with $\mu \to e\nu\nu$ decays.  Data from the
 $\Upsilon(4S)$ resonance, and off resonance data are used
 as calibration and control samples. 

\begin{figure}[htb]
\includegraphics*[width=3.2in]{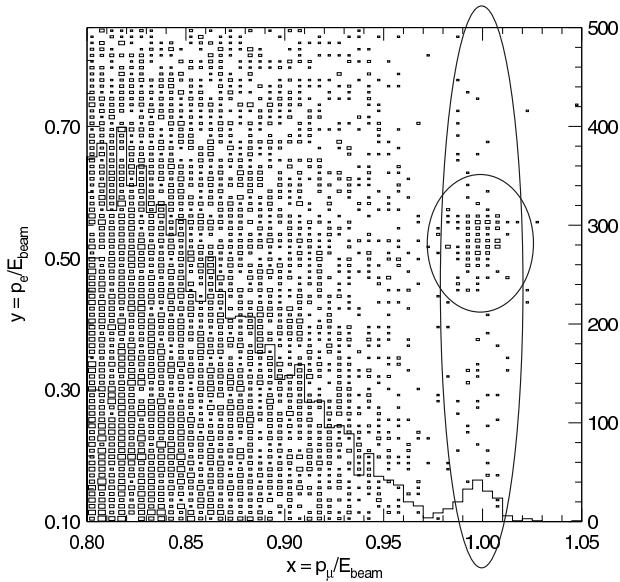}
 \caption {
  Beam energy scaled electron  versus
 the scaled muon momentum in $\Upsilon(4S)$ data.
 The left hand side is dominated by $\tau$ pair
 decays. The signal region is a vertical strip near
 $p_\mu/E_{Beam} \approx 0.97$. The concentration
 of events near  $p_\mu/E_{Beam} =1 $ and $p_e/E_{Beam}=0.5$
 is from $\mu\mu\gamma$ backgrounds. Also at  $p_\mu/E_{Beam} =1$
 are events from muon decays to electrons in $\mu\mu$ production. }
\label{fig:pevspmu}
\end{figure}

 Fig~\ref{fig:pevspmu} shows the beam energy scaled electron momentum spectrum versus
 the scaled muon momentum spectrum, in $\Upsilon(4S)$ data where
 no signal is expected, since even $\Upsilon(4S)\to \tau \tau$ is too
 small to be observed. 

\begin{figure}[htb]
\includegraphics*[width=3.2in]{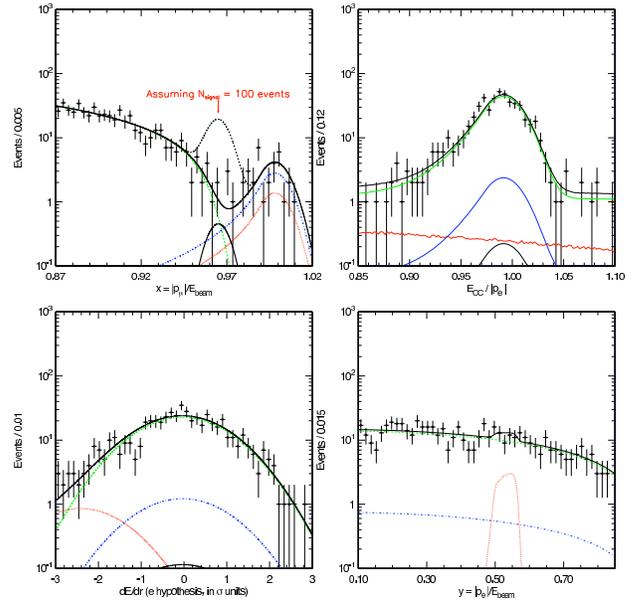}
 \caption {
Likelihood distributions for the scaled muon momentum, E/p, dE/dx and
 scaled momentum for the electron candidate.  Note the curve indicating
 what 100 signal events would look like.
 }
\label{fig:likelihood}
\end{figure}

 Fig~\ref{fig:likelihood} shows the data distributions for the fit quantities at
 the $\Upsilon(1S)$. Note the clear absence of a signal, especially
 in the muon momentum spectrum. 

 Preliminary analysis results are summarized in Table~\ref{tab:mutau}.
 The largest systematic uncertainties are  from PDF shapes
 and their correlations. This study represents the first
 limits on $B(\Upsilon \to \tau\mu)$. These limits set
 a lower limit of $\approx 1 $ TeV on the scale of new
 physics, assuming a strong coupling.

\begin{table*}[htb]
  \vspace{0.4cm}
 \begin{center}
 \begin{tabular}{ || c |  c |  c |  c || }
 \hline 
Resonance & $\Upsilon(1S)$ & $\Upsilon(2S)$ &  $\Upsilon(3S)$ \\ \hline 
Efficiency & $8.9\%$ &  $8.9\%$ &  $8.9\%$ \\  \hline 
Events & $< 10.0 $ & $< 10.7 $ &$< 8.5 $ \\ \hline 
$B(\Upsilon \to \tau\mu)$ ($10^{-6}$)& 
         $ < 6.2 $ &   $ < 25 $ &   $ < 22 $ \\ \hline 
$B(\tau\mu)/B( \mu\mu)$ &
         $ < 0.023 \%$ &  $ < 0.17 \%$ & $ < 0.13 \%$ \\
\hline
\end{tabular}
\caption{ Preliminary results for the LFV search. Displayed are the
 signal MC efficiencies, raw event counts, resulting 
 branching fractions, and the ratio $B(\Upsilon\to\tau\mu)/
 B( \Upsilon\to\mu\mu)$. All limits are $90 \%$ confidence
 level upper limits.}
\label{tab:mutau} 
\end{center}
\end{table*}

\section{ Lepton Universality in $\Upsilon \to \mu\mu, \tau\tau$ }

 In the Standard Model,  the
 couplings between leptons and gauge bosons are independent of the lepton flavor, so
  the branching fractions for the decay $\Upsilon(nS)\to {\it l^+}{\it 
l^-} $ should be
   independent of the flavor of the lepton ${\it l}$, 
except for negligible final state lepton mass effects.
Any deviation from unity for
    the ratio of branching fractions
     $ \myratio = B(\Upsilon(nS) \to \tau\tau)/B(\Upsilon(nS) 
\to \mu\mu) $ 
 would indicate the presence of new physics.
The ratio $\myratio$ is sensitive to the mechanism proposed in \cite{miguel}, 
in which a low mass CP-odd Higgs boson, $A^0$,
mediates the  decay chain $\Upsilon(1S) \to \eta_b \gamma, \eta_b \to A^0 \to
 \tau\tau$.

CLEO has recently measured the partial width, $\Gamma_{ee}$, from
 $e^+e^- \to \Upsilon(nS)$~\cite{pivarski}, 
 $n=1,2,3$, as well as the branching fraction for $\Upsilon(nS) 
\to \mu^+\mu^-$~\cite{idanko}.
  This analysis complements these measurements by measuring 
  $\myratio$ directly, and scales this result to obtain 
$B(\Upsilon(nS) \to \tau\tau)$.
An upper limit on the product branching fraction 
 $B( \Upsilon(1S) \to \eta_b \gamma) B( \eta_b \to A^0 \to \tau\tau )$ 
is extracted.

The analysis technique, similar to that  in \cite{idanko}, isolates 
the $\Upsilon \to \mu\mu , \tau\tau$ signals by subtracting a luminosity 
and beam energy weighted number of events observed in off-resonance data 
from the number observed in 
on-resonance data, and, after further background correction, attributes the 
remaining signal to $\Upsilon$ decays to leptons.
Selection criteria are developed 
to isolate $\mu\mu$ and $\tau\tau$ final states using a subset 
of the data acquired near the $\Upsilon(4S)$. Another subset of 
 $\Upsilon(4S)$ data is used to verify
that subtracting the scaled off-resonance data from the  
on-resonance data produces no signal for $\Upsilon(4S)\to \tau\tau, \mu\mu$,
indicating  that non-$\Upsilon$ backgrounds are suppressed by the subtraction. 
As a further crosscheck, the  off-resonance
 production cross-sections for $\tau\tau$ and $\mu\mu$ are verified to
 agree with theoretical expectations.

The 
final states chosen for both the $\Upsilon \to \mu\mu$ and $\Upsilon \to 
\tau\tau$ decays are required to have exactly two good quality charged 
tracks of opposite charge. 

Selection criteria for the $\mu\mu$ final state closely parallel those 
of \cite{idanko}, requiring  tracks with momenta scaled to $\ebeam$  between 
0.7 and 1.15, of which at least one is positively identified  as a muon. 
The energy deposited by a particle in a calorimeter shower, $E_{\mathrm{CC}}$,
is required to satisfy $100 \,\mathrm{MeV} < E_{\mathrm{CC}} < 600 
\,\mathrm{MeV}$.
No more than one  shower unassociated with a track and with energy 
above 1\% of $\ebeam$ is allowed. 

At least two neutrinos from final states of $e^+e^- \to \tau\tau$ escape detection.
Thanks to CLEO-III's hermeticity, the following criteria select such events despite
 the energy carried away by the unreconstructed neutrinos.
 The total charged track momentum transverse to the beam direction must be greater
than 10\% of $\ebeam$, and the total charged track momentum must point into the barrel region
 of the detector where tracking and calorimetry are optimal. Events with 
collinear tracks are eliminated. Tracks are required to have momenta greater 
than 10\% of $\ebeam$ to ensure that they are well-reconstructed, and, to 
minimize pollution from two-particle final states, they are required to have 
momenta less than $90\%$ of $\ebeam$.  The total observed energy  due to 
charged and neutral particles in the calorimeter is similarly required to be 
between $20\%$ and $90\%$ of the total center-of-mass energy. To reduce 
overlap confusion between neutral and charged particles, a shower's energy 
scaled to its associated track's momentum must be less than 1.1.

Final states are further exclusively 
divided according to the results of particle identification into $(e,e)$, 
$(\mu,e)$, $(e,\mu)$, $(\mu,\mu)$, $(e,X)$, $(\mu,X)$, $(X,X)$ sub-samples, 
with
  particles listed in descending momentum order.
 The first (second) particle listed
 is referred to as the tag (signal). The sub-samples are used to
crosscheck consistency of results across decay modes.
 Lepton identification requires a track momentum greater than $500 \, \mathrm{MeV}$
 to ensure that the track intersects the calorimeter.
Electrons are identified by requiring that  $0.85 < E_{\mathrm{CC}}/P < 1.10$, where $P$
 is the track momentum, and that the specific ionization along the track's 
path in the drift 
chamber be consistent with the  expectation for an electron.
A muon candidate in $\tau$ decays is a charged track which is not identified as an 
electron, having momentum above $2 \,\mathrm{GeV}$ ($1.5 \,\mathrm{GeV}$) 
for a tag (signal) track
  and confined to the central barrel 
where beam related background is a minimum.
 Furthermore, the energy deposited in the calorimeter for this 
track must be between $100$ and $600 \,\mathrm{MeV}$, and the particle must 
penetrate at least three interaction lengths into the muon detector.  
Particles identified as neither e nor $\mu$ are designated $X$, and are 
a mixture of hadrons and unidentified leptons.

The decay products of $\tau$ pairs from $\Upsilon$ decays 
tend to be separated into distinct hemispheres.
Since the photon spectrum expected in $\tau$ decays depends 
on the identity of the charged particle, calorimeter showers are 
assigned to either the tag or signal hemisphere according to 
their proximity to the tag side track direction.
 For each $\tau$ decay mode pair,
the number of unmatched showers in the calorimeter as well as the total
energy of these showers on each side of the decay are used as selection
criteria.

For the 
$(e,e)$ and $(\mu,\mu)$ 
modes, the sum of the magnitudes of the tag and signal track momenta must be less 
than $1.5 \ebeam $, reducing the contamination from radiative dilepton
 events.
To reduce backgrounds from $e^+e^- \to {\it l}^+  {\it l}^- 
\gamma \gamma$ and 
$e^+e^- \to e^+e^- {\it l}^+ {\it l}^-$ in the $(e,e)$, $(\mu,\mu)$, 
$(X,X)$ categories, the minimum polar angle
 of any unseen particles, deduced from energy-momentum conservation,
 is required to point into the barrel region, where 
 calorimetry cuts will ensure rejection.

Potential backgrounds due to cosmic rays are accounted for 
as in \cite{idanko}. In all cases these were negligible. 

\begin{figure}
\includegraphics*[width=3.2in]{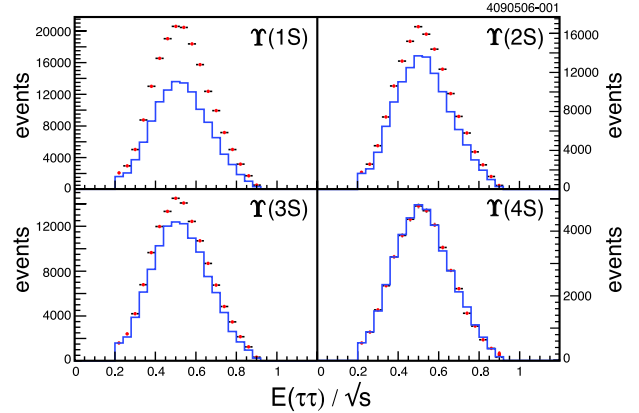}
\caption{ Total energy distributions, scaled to center of mass energy, 
$E_{\tau\tau}/\roots$
for the $\tau\tau$ final states
 at the $\Upsilon(nS), n=1,2,3,4$ on (points) and scaled off (histograms) resonance. 
 The excess of
 on-resonance relative to scaled off-resonance data 
 is attributed to $\Upsilon$ decays for the lower resonances, while the agreement
 at the $\Upsilon(4S)$ tests the validity of the subtraction. 
}
\label{fig:onoff}
\end{figure}

Figure~\ref{fig:onoff} shows the superimposed on-resonance  and scaled 
off-resonance 
total energy distributions
 for the $\tau\tau$ sample for all resonances. The scale factor is 
$S =  ( {\cal L}_{\mathrm{On}}/{\cal L}_{\mathrm{Off}})
            ( s_{\mathrm{Off}}/ s_{\mathrm{On}}) \delta_{\mathrm{interf}}$, 
where
 ${\cal L}$ and $s$ are the data luminosity and squared center of mass 
 collision energies on and off the resonances, and $\delta_{\mathrm{interf}}$
 is an interference correction.
  The luminosity 
 is derived from the  process 
$e^+e^-\to\gamma\gamma$ \cite{lumi},
 which does not suffer backgrounds from direct $\Upsilon$ decays. The interference
  correction $\delta_{\mathrm{interf}}$ accounts for the small interference between
   the process $e^+e^- \to {\it l l}$ and $e^+e^- \to \Upsilon \to {\it l l}$ and is estimated~\cite{idanko} 
   to be 0.984 (0.961, 0.982) at the $\Upsilon(1S)$ $(2S,\, 3S)$ and negligible for the $\Upsilon(4S)$. Note
    that the interference largely cancels in the ratios considered in this work.
 The agreement
 of the distributions for the $\Upsilon(4S)$ validates the 
subtraction technique, and also highlights the
absence of any process  whose cross-section does not vary as $1/s$.
This agreement 
 extends to the individual sub-samples.

The ratio of measured relative lepton pair 
 production cross-sections, $\myratiooff = \sigma_{\tau\tau}/\sigma_{\mu\mu} $,
 with respect
 to that theoretically expected  at the off-resonance points, 
 $\myratiooffexp =  \sigma_{\tau\tau}^{\mathrm{theory}}/\sigma_{\mu\mu}^{\mathrm{theory}} $
  is 
 $\myratiooff/\myratiooffexp = 0.96\pm 0.03$ ($0.97 \pm 0.03 $, 
 $0.97 \pm 0.03 $ , $1.00 \pm 0.03 $) below the $\Upsilon(1S)$ 
 ($2S$, $3S$, $4S$) for the sum of all
 $\tau$ decay mode pairs, with statistical and systematic uncertainties 
 added in quadrature.
 The expectation
 $\myratiooffexp= 0.83 \pm 0.02 \,\mathrm{(syst)} $~\footnote{
 This expectation is lower than 1 because of the larger phase space available for
  initial state radiation production of $\mu\mu$ relative to $\tau\tau$ final states.}, which is found
 to be numerically independent of the particular resonance considered,
 and the
 reconstruction efficiencies are derived 
from the FPair~\cite{FPair} and 
Koralb/Tauola~\cite{koralb,tauola,photos} Monte Carlos. Backgrounds were 
corrected by using $e^+e^- \to q \overline{q} (q=u,d,c,s)$ Monte 
Carlo simulations~\cite{QQ}~\cite{Jetset}~\cite{photos}~\cite{Geant}. 
 The scatter in the central values of $\myratiooff/\myratiooffexp$ indicates
 that systematic uncertainties are small.

The reconstruction efficiency for observing $\Upsilon \to \mu\mu$ is derived 
from the CLEO  {\textsc GEANT}-based simulation~\cite{QQ}~\cite{Jetset}~\cite{photos}~\cite{Geant},
 as shown in Table~\ref{tab:NumberOfEvents}.
This efficiency is found to be constant
across the resonances.  
 
 The reconstruction efficiency for observing $\Upsilon \to \tau\tau$ is 
derived using the Koralb/Tauola event generator integrated 
into the Monte Carlo simulation. Although this generator
  models  the process $e^+e^- \to \gamma^* \to \tau^+\tau^-$, the 
quantum numbers of the $\Upsilon$ and $\gamma$ are the same so
it can be used
  as long as initial state radiation (ISR) effects are not included. 
This efficiency is also found to be consistent across all
 resonances within any given $\tau\tau$ decay channel.

  Results of the subtraction are summarized in Table~\ref{tab:NumberOfEvents},
 showing 
the first observation of $\Upsilon(3S)  \to \tau\tau$.

  \begin{table*}[htb]
  \vspace{0.4cm}
\renewcommand{\tabcolsep}{2pc} 

 \begin{center}
 \begin{tabular}{ c  c  c  c }
 \hline 
 	&
 $\Upsilon(1S)$ \,\,&
 $\Upsilon(2S)$ \,\,&
 $\Upsilon(3S)$ \\ \hline
 \rule[-0.5mm]{0mm}{4mm}
$ \tilde{N} ( \mu\mu)$ ($10^3$)&
 $345 \pm 7$ \,\,&
 $121\pm 7$ \,\,&
 $82 \pm 7$ \\ 
 $ \epsilon(\mu\mu)$ ($\%$) &
  $65.4 \pm 1.2 $ \,\,&
  $65.0 \pm 1.1 $ \,\,&
  $65.1 \pm 1.2 $ \\ 
  $N(\Upsilon \to \mu\mu) $ ($10^3$) &
   $527 \pm 15 $ \,\,& 
   $185 \pm 11 $ \,\,&
    $ 126 \pm 11 $ \\ \hline 
 \rule[-0.5mm]{0mm}{4mm}
 $ \tilde{N} (\tau\tau) $ ($10^3$)&
 $60.1 \pm 1.5 $ \,\,&
 $21.8\pm 1.5 $ \,\,&
 $14.8 \pm 1.5 $ \\ 
 $ \epsilon(\tau\tau)$ ($\%$) &
  $11.2 \pm 0.1  $ \,\,&
  $11.3 \pm 0.1 $ \,\,&
  $11.1 \pm 0.1 $ \\ 
  $N(\Upsilon \to \tau\tau) $ ($10^3$) &
   $537 \pm 14 $ \,\,& 
   $193 \pm 12 $ \,\,&
    $ 132 \pm 13 $ \\ \hline 
\end{tabular}
\caption{ Summary of reconstructed events for $\Upsilon \to \mu\mu$ (top)
 and  $\Upsilon \to \tau\tau$ (bottom). 
 Shown are the number of events  ($\tilde{N}_{\it l l}$) after subtraction of backgrounds estimated
  from scaled off-resonance data and $\Upsilon$ feed-through estimated from the
   Monte Carlo simulation,
 the signal efficiency ($\epsilon({\it l l})$), and 
 the total efficiency corrected number of signal events $N(\Upsilon \to {\it ll})=\tilde{N}_{\it l l}/\epsilon({\it l l} )$.
 The $\tau\tau$ events are summed over all decay modes of the $\tau$.
 Uncertainties included in this table include data and Monte Carlo statistical 
uncertainties, uncertainties on backgrounds, and detector modeling (included only
 for $\epsilon(\mu\mu)$ to avoid double counting in the final ratio).}
\label{tab:NumberOfEvents} 
\end{center}
\end{table*}

 Backgrounds resulting from  
 cascade decays within the $b\overline{b}$ system to ${\it l l}$ are
  estimated using the Monte Carlo simulation, 
  with branching fractions scaled to the values measured in this study.
 Cascade backgrounds with
 non  $\mu\mu$ and non $\tau\tau$ final states are estimated directly from 
 the  Monte Carlo simulation.

\begin{figure}
\includegraphics*[width=3.2in]{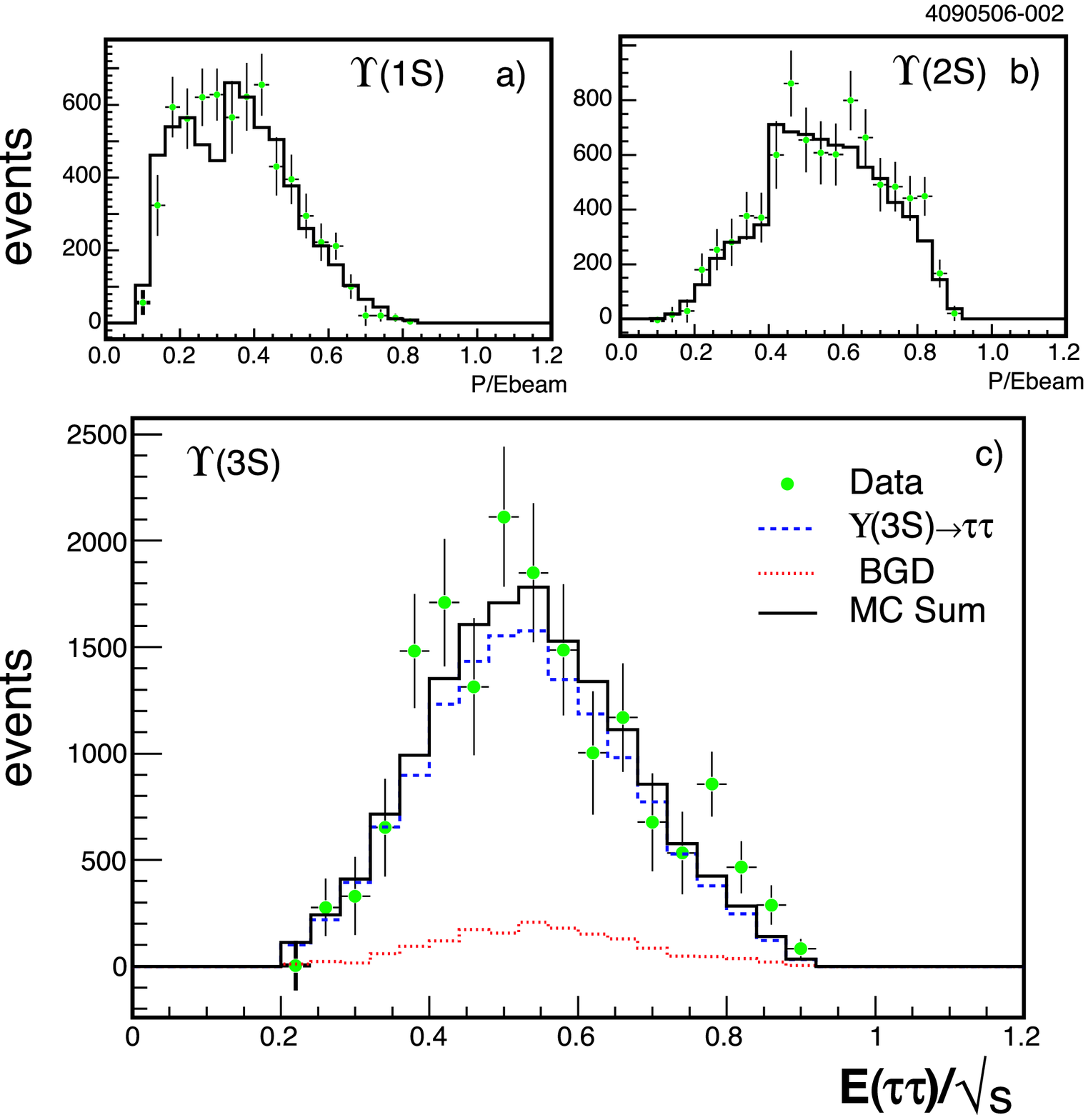}
\caption{ Distributions for the $\tau\tau$ final states
 at the $\Upsilon(nS), n=1,2,3$ after subtraction of $S$-scaled off-resonance
 data. Distribution a) shows $P_{\mathrm{sig}}/\ebeam$
 in $\Upsilon(1S)$ decays for the sum of $\tau$ decay 
 modes including exactly
 two identified leptons.  Distribution b) shows 
 $P_{\mathrm{tag}}/\ebeam$
 in $\Upsilon(2S)$ decays for the sum of
 $\tau$ decay modes including exactly one identified lepton. 
 Distribution c) shows 
 $E_{\tau\tau}/\roots$  for $\Upsilon(3S)$ for
 the sum of all $\tau$ decay modes. 
 In all cases, the solid line shows the expected total signal and background
 distributions, assuming lepton universality. In distribution c), the signal and
  total background distributions are explicitly displayed.
  Data uncertainties shown are purely statistical.
}
\label{fig:signal}
\end{figure}

Figure~\ref{fig:signal} displays the off-resonance subtracted data, 
superimposed on Monte Carlo expectations.
The distributions shown are a sampling of $\tau$ decay modes for
 the momenta of the signal and tag tracks, as well as the total reconstructed
 energy. In all cases the Monte Carlo expectations are consistent with the
 data assuming lepton universality and branching fractions as measured in
 ~\cite{idanko}. The agreement across the various
 kinematic quantities indicates that backgrounds are well controlled.

\begin{figure}
\includegraphics*[width=3.0in]{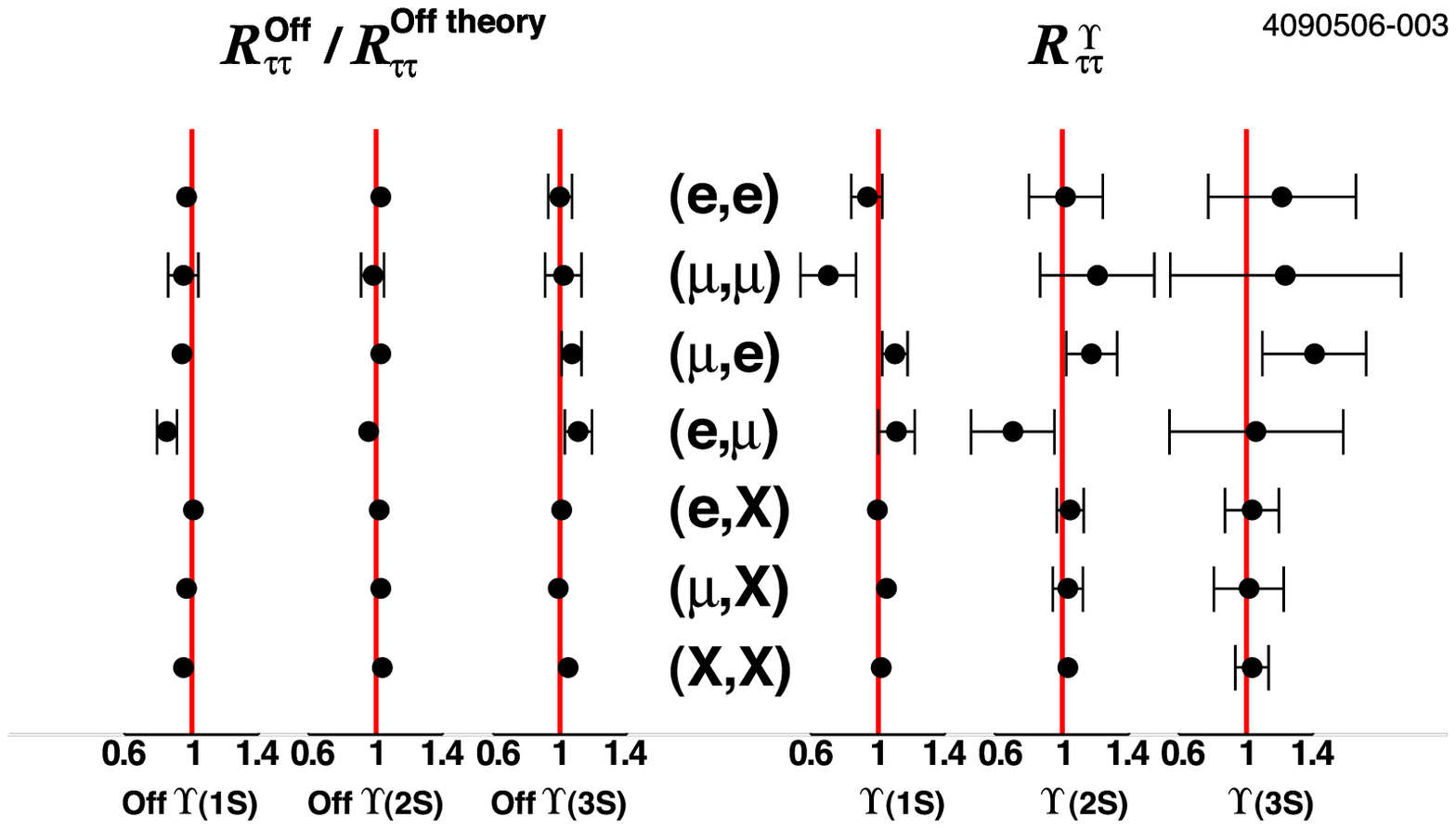}
\caption{ Breakdown by mode of off- and on-resonance data 
 at the different resonances.
 On the left, the ratio of the production cross section for 
 $e^+e^- \to {\it l}{\it l} ({\it l} = \tau,\mu)$,
 relative to its expectation,
 is plotted for data taken below the $\Upsilon$ for 
 each $\tau $ decay mode pair. On the right, the ratio of branching
 fractions for the process $\Upsilon\to {\it l}{\it l } ({\it l} = \tau,\mu)$
 is displayed.
The line centered at 1 in each case represents the Standard Model
expectation. Errors
 shown are statistical.
}
\label{fig:breakdown}
\end{figure}

Figure~\ref{fig:breakdown} shows the agreement across all $\tau\tau$ 
 sub-samples of the ratio of off-resonance cross sections for $\tau\tau$ and
 $\mu\mu$ production, relative to expectation, as well as the ratio of 
branching fractions for each
 of these decay modes at the different $\Upsilon$ resonances.
 The agreement across $\tau\tau$ sub-samples both on and off the resonances is
 again an indication that backgrounds are small and well estimated.

The ratio of branching fractions and final branching fractions are listed in 
 Table~\ref{tab:FinalResults}. These results show that lepton universality
 is respected  in $\Upsilon$ decay 
 within the $\approx 10\%$ measurement uncertainties.

  \begin{table*}[htb]
  \vspace{0.4cm}
 \begin{center}
\renewcommand{\tabcolsep}{2pc} 
 \begin{tabular}{ c  c   c }
 \hline
 \rule[-0.5mm]{0mm}{4mm}
      \,\,\,\,   &
 $\myratio$ \,\,\,\, &
 $B(\Upsilon \to \tau \tau $ ) ($\%$) 	\\  \hline
 $\Upsilon(1S)$ \,\,\,\, &
 $1.02 \pm 0.02 \pm 0.05$ \,\,\,\, &
 $2.54 \pm 0.04 \pm 0.12 $ \\
 $\Upsilon(2S)$ \,\,\,\,  &
  $1.04 \pm 0.04 \pm 0.05$ \,\,\,\, &
 $2.11 \pm 0.07 \pm 0.13 $ \\
 $\Upsilon(3S)$ \,\,\,\, &
   $1.07 \pm 0.08 \pm 0.05$ \,\,\,\, & 
   $2.55 \pm 0.19 \pm 0.15 $  \\ \hline
\end{tabular}
\caption{ 
 Final results on the ratio of branching fractions to $\tau\tau$ and
 $\mu\mu$ final states, and the absolute branching fraction for
 $\Upsilon \to \tau\tau$. Included are both statistical and systematic
 uncertainties, as detailed in the text.
\label{tab:FinalResults} }
\end{center}
\end{table*}

 Systematic uncertainties, summarized in Table~\ref{tab:syst}, are estimated for
 the ratio of branching fractions, 
and for the absolute branching fraction. 
  The ratio of branching fractions is insensitive to some common systematic
 uncertainties. 
 For instance, the uncertainty on $\myratio$ due to a conservative 
 $1\%$ variation
 in the scale factor $S$, 
 as determined using the process 
 $e^+e^-\to\mu\mu$ near the $\Upsilon(4S)$ resonance,
 is found to be  $0.4\%$ or less. 
 
 \begin{table}[htb]
 \vspace{0.4cm}
 \begin{center}
 \begin{tabular}{ c  c }
 \hline
 Source        &
 $\sigma_{syst} $ ($\%$) \\ \hline
 $S $ & $0.2$ / $0.4$ / $0.3$ \\ 
 Background & $0.1$ / $2.4$ / $1.3$ \\ \hline 
 $\tau$, $\mu$ Selection & $2.9$ \\ 
 $\Upsilon \to \mu\mu$ Model & $2.0$ \\ 
 $\Upsilon \to \tau\tau$ Model & $2.0$ \\ 
 Detector Model & $1.7$ \\ 
 MC Statistics & $1.9$ / $1.0$ / $1.0$ \\ \hline 
 \rule[-0.5mm]{0mm}{4mm}
 $\sigma(\myratio)/\myratio$ & $4.8$ / $4.4$ / $4.6$ \\ \hline 
  $\sigma(B_{\tau\tau})/B_{\tau\tau}$  & $4.0$ / $3.8$ / $3.9$ \\ \hline
\end{tabular}
\caption{ 
Summary of systematic uncertainties.
 The entry $\sigma(\myratio)/\myratio$  indicates the relative uncertainty on 
$\myratio$, while the $\sigma(B_{\tau\tau})/B_{\tau\tau}$ entry 
 indicates  uncertainties
 specific to $\tau$ decay modes used in addition to
 those in \cite{idanko} to obtain $B(\Upsilon \to \tau\tau)$.
 The uncertainty on $S$ and the background are included in the statistical
  uncertainty only. Lines with three entries indicate the contribution from the
 $\Upsilon(1S)$,  $\Upsilon(2S)$ and  $\Upsilon(3S)$, respectively.
 \label{tab:syst} }
\end{center}
\end{table}

 Most systematic uncertainties due specifically to ${\it l}{\it l}$ selection
are 
 derived by a variation of the selection criteria
 over reasonable ranges in the $\Upsilon(1S)$ sample, which has
 the lowest energy released in its decay.
The most significant of these are due to 
momentum selection ($1.3\%$), calorimeter energy selection ($1.1\%$) and 
angular selection ($1.1\%$).
  The systematic uncertainty due to modeling of the trigger, also included in the
${\it l}{\it l}$ selection,  is estimated by using a
 loose pre-scaled tracking trigger and comparing it to the more sophisticated
 triggers used in this analysis. This variation leads to a systematic uncertainty
 estimate of $1.6\%$.

Backgrounds are assumed to be due solely to $\Upsilon$ decays. 
As in \cite{idanko}, these 
were chiefly due to cascade decays to lower resonances, and are estimated
 to be $2.5\%$ ($15\%$, $11 \%$) of the $\tau\tau$ sample at the $\Upsilon(1S) ( 2S, 3S)$, with
 an estimated uncertainty contribution to  $\myratio$ of $0.1\%$ ($2.4\%$, $1.3\%$).
 
The uncertainty due to detector modeling
 in \cite{idanko} was estimated to be $1.7\%$: this 
value is used
   here conservatively for the systematic uncertainty on the ratio. 
  
  The modeling of the physics in $\Upsilon(1S) \to \mu\mu$,
 obtained by varying the decay model for
 $\Upsilon \to \mu\mu$  between the Monte Carlo simulation
  and Koralb with ISR simulation turned off,
contributes an uncertainty $2\%$.
This uncertainty is  consistent with the variation 
in the product of the off-resonance cross section and reconstruction 
efficiency using the FPair, 
     Koralb, and Babayaga Monte Carlo simulations, and is thus likely conservative, 
as direct $\mu\mu$ production from
the $\Upsilon$ at the peak involves much lower energy final state
 photons than off resonance production.
An uncorrelated uncertainty of 
$2\%$ for modeling of $\Upsilon \to \tau\tau$ is assumed,
consistent with the uncertainty on the off-resonance
production cross section derived in previous analyses, and is again 
 conservative as 
     on-resonance production of $\tau\tau$ final states involves fewer 
photons than direct
      continuum production. 
To test the sensitivity to ISR simulation, the reconstruction efficiency for events
with no ISR simulation is compared to that for events generated
with ISR simulation turned on
and re-weighted according to the relative value of the $\Upsilon$ line
 shape at the $\tau$ pair mass. 
 These two efficiencies agree to within $0.8\%$ of their central value.

 The mechanism described in \cite{miguel} could induce
 a value of $\myratio$ not equal to one. By assuming that the
 mass of the $\eta_b(1S)$ is $100 \,\mathrm{MeV/c}^2$ below the $\Upsilon(1S)$ 
mass,
consistent with the largest value in \cite{Rosner},
 the value quoted for $\myratio (1S)$ can be translated into an upper limit
 on the combined branching fraction of $B(\Upsilon(1S) \to \eta_b \gamma)
 B(\eta_b \to A^0  \to \tau \tau) < 0.27\% $ at $95\%$ confidence level.
Since the transition photon is not explicitly reconstructed,
this limit is valid for all $\eta_b$ that approximately
 satisfy 
 $M(\Upsilon(1S)) - M(\eta_b) + \Gamma(\eta_b)) < 
{\cal O } (100 \,\mathrm{MeV/}c^2)$.

 In summary, using the full sample of on-resonance $\Upsilon(nS)$, $n=1,2,3$, collected
 at the CLEO-III detector, we have made the first observation of the
 decay $\Upsilon(3S) \to \tau\tau$. We have also reported the ratio
 of branching fractions of $\Upsilon$ decays to $\tau\tau$ and $\mu\mu$
 final states, and find these to be consistent with expectations from the
 Standard Model. These ratios have been combined with results from \cite{idanko}
 to provide absolute branching fractions for the process 
 $\Upsilon \to \tau\tau$, shown in  Table~\ref{tab:FinalResults},
 resulting in the most precise single measurement of 
 $B(\Upsilon(1S)\to\tau\tau)$~\cite{pdg}, 
 a much improved value of $B(\Upsilon(2S)\to\tau\tau)$ and
 a first measurement of $B(\Upsilon(3S) \to \tau\tau)$.
  The ratio of branching fractions for $\tau\tau$ and $\mu\mu$ final states
 has also been used to set a limit on a possible Higgs mediated decay window.

We gratefully acknowledge the effort of the CESR staff in providing us with
 excellent luminosity and running conditions. 
We also thank M.A.~Sanchis-Lozano for many illuminating discussions.
This work was supported by 
the A.P.~Sloan Foundation,
the National Science Foundation,
the U.S. Department of Energy, and
the Natural Sciences and Engineering Research Council of Canada.

%


\end{document}